# Ultra-short pulse transmission in a passive white light cavity based on chirped Bragg gratings

**Alexander Pavlov[1], Mike Lee[2] and Saito Fujimoto[3]**

[1]*Moscow State Engineering Physics Institute, Moscow, Russia*

[2]*University of Technology Sydney, Sydney, Australia*

[3]*Kagoshima University, Kagoshima, Japan.*

*Corresponding author: Alexander.Pavlov1970@gmail.com

**Abstract:** It is widely accepted that Bragg reflection from linearly chirped Bragg gratings (LCBGs) can compensate for a chromatic dispersion by reflecting different wavelengths at different location along the axis of the gratings. In this paper, we studied the possibility of making use of such a dispersion-compensating property to construct a white light cavity (WLC). A pair of LCBGs is suggested as the reflectors of the cavity. The analytical and numerical analysis shows that the reflection by a LCBG does not solely occur at the position where Bragg condition is reached. The accumulated effect of multiple scatterings at different locations inside the LCBG produces a positive group delay, preventing the WLC to be realized with simple LCBG.

**References and links**

1. R. H. Rinkleff, and A. Wicht, "The concept of white light cavities using atomic phase coherence," Phys. Scr. T. 118,85–88 (2005).
2. A. Rocco, A. Wicht, R.-H. Rinkleff, and K. Danzmann, "Anomalous dispersion of transparent atomic two- and three-level ensembles," Phys. Rev. A, 66, 053804 (2002).
3. A. Wicht, K. Danzmann, M. Fleischhauer, M. Scully, G. Miller, and R. H. Rinkleff, "White-light cavities, atomic phase coherence, and gravitational wave detectors," Opt. Commun. 134(1-6), 431–439 (1997).
4. M. Scalora, N. Mattiucci, G. D'Aguanno, M. C. Larciprete, and M. J. Bloemer, "Nonlinear pulse propagation in one-dimensional metal-dielectric multilayer stacks: Ultrawide bandwidth optical limiting," Phys. Rev. E 73, 016603 (2006).
5. X. Liu, J. W. Haus and S.M. Shahriar, "Modulation instability for a relaxational Kerr medium," Opt. Comm. 281, 2907-2912 (2008).
6. P. Tran, "All-optical switching with a nonlinear chiral photonic bandgap structure," J. Opt. Soc. Am. B 16, 70-73 (1999)..
7. Xue Liu, Joseph W. Haus and M. S. Shahriar, "Optical limiting in a periodic materials with relaxational nonlinearity," Opt. Exp. 17, 2696-2706 (2009).
8. Yum, H., Liu, X., Jang, Y. J., Kim, M. E., & Shahriar, S. M. "Pulse delay via tunable white light cavities using fiber-optic resonators," Journal of Lightwave Technology, 29, 2698-2705 (2011).
9. R. Fleischhaker, and J. Evers, "Four wave mixing enhanced white-light cavity," Phys. Rev. A 78(5), 051802 (2007)
10. B. J. Meers, "Recycling in laser-interferometric gravitational-wave detectors," Phys. Rev. D 38, 2317 (1988).

11. G. Heinzel, K.A. Strain, J. Mizuno, K. D. Skeldon, B. Willke, W. Winkler, R. Schilling,1 A. Rüdiger,1 and K. Danzmann, "Experimental demonstration of a suspended dual recycling interferometer for gravitational wave detection" Phys. Rev. Lett. 81 No. 25 5493 (1998).
12. M. B. Gray, A. J. Stevenson, H.-A. Bachor, and D. E. McClelland, "Broadband and tuned signal recycling with a simple Michelson interferometer,"Appl. Opt. 37, 5886 (1998).
13. S. Wise, G. Mueller, D. Reitze, D. B. Tanner, and B. F.Whiting, "Linewidth-broadband Fabry-Perot cavities within future gravitational wave detectors," Classical Quantum Gravity 21, S1031 (2004).
14. S. Wise, V. Quetschke, A.J. Deshpande, G. Mueller, D.H. Reitze, D.B. Tanner and B.F. Whiting, "Phase effects in the diffraction of light: beyond the grating equation," Phys. Rev. Lett. 95, 013901 (2005).
15. O.V. Belai, E.V. Podivilov, D.A. Shapiro, "Group delay in Bragg grating with linear chirp," Opt. Comm., 266, 512-520 (2006).
16. P. Yeh, 'Optical Waves in Layered Media', Wiley (1991).J. S. Shirk, R. G. S. Pong, F. J. Bartoli, and A. W. Snow, "Optical limiter using a lead phthalocyanine," Appl. Phys. Lett. **63**, 1880-1882 (1993).

## 1. Introduction

White light cavity (WLC) has a broader linewidth without loss of a build-up factor than an ordinary cavity with the same finesse[1,2]. Such an enhanced linewidth is an essential property to apply WLC to optical detection, sensing and communication: For example WLC increases the sensitivity enough to detect the extremely weak side band signal produced by gravity waves without restricting the detection bandwidth [3]. For a data buffer system, the transmission over a broad spectral range to encompass the data pulse spectrum has been studied through many mechanisms including nonlinear medium[4-8]. It shows an enhanced delay time bandwidth product (DBP) overcoming constraints encountered by conventional buffer systems. WLC effect in a laser cavity was proposed for hypersensitive rotation sensing wherein the sensitivity of the lasing frequency to displacement was enhanced on the order of ~$10^5$ higher than a conventional laser [8-10].

In WLC, the frequency dependent phase shift due to propagation delay is cancelled by tailoring a dispersion profile of the intracavity medium [11-13] such that WLC condition is achieved: $n_g = 1 - L/\ell$ where $L$ is the cavity length, and $n_g$ and $\ell$ are the group index and the length of intracavity medium, respectively. In previous implementations, $n_g$ is controlled by coupling a weak probe to a strong pump in non-linear media. However, such a probe-pump interaction scheme is not applicable when we need to use a high power probe, for example, ~40 Watt probe beam is used in the Advanced LIGO interferometer and then the pump would have to be even stronger to induce dispersion. Current material technology cannot provide a non-linear medium which holds such high power beams. A passive approach to the enhancement of the bandwidth of LIGO like interferometer was attempted by using two gratings placed in parallel; however, if one consider the geometrical optical path arising from the wavelength-dependent diffraction angle as well as the additional phase change associated with the spatial phase modulation of the gratings, it is then impossible to make the variation of the phase with respect to frequency become zero. The essence of this constraint originates from the constant grating period. In ref.[13], if the grating period is a function of frequency i.e. $g \equiv g(\omega)$ rather than the constant, then the phase variation would become $d\Phi/d\omega = \beta L(\omega)/c - D\tan[\partial g/\partial \omega]$ implying that $d\Phi/d\omega$ could be zero with the appropriate choice of $dg/d\omega$.

in this paper, we studied the possibility of making use of such a dispersion-compensating property to create a white light cavity (WLC). A pair of LCBGs is suggested as the reflectors of the cavity. The analytical and numerical analysis show that the reflection by a LCBG does not solely occurs at the Bragg matching point. The accumulated effect of multiple scatterings at different location inside the LCBG produces a positive group delay, which is contradictory to conventional notion about the LCBG.

## 2. White Light Cavity (WLC) condition

First, it is necessary to understand the condition needed in order to construct a passive WLC. As show in Fig.1, the basic scheme of the WLC is a variation of a regular Fabre-Perot interferometer with additional phase generated by the reflectors.

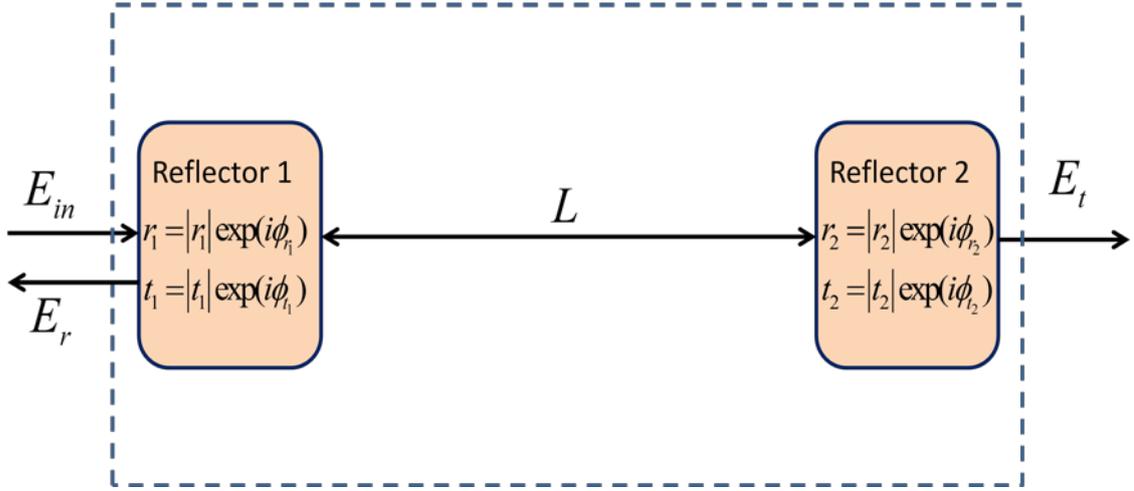

Fig. 1. Schematic illustration of WLC: two reflectors are separated by L to form a modified FP cavity.

The cavity has a length $L$ between the surfaces of the reflectors and a mean refractive index $n_0$. The field reflection coefficient and transmission coefficient of the two reflectors are denoted as $r_i$ and $t_i$ ($i=1,2$), respectively. Note that $r_i = |r_i| exp(j\phi_{r_i})$ and $t_i = |t_i| exp(j\phi_{t_i})$ are complex numbers with the phase shift $\phi_{r_i}$ and $\phi_{t_i}$ upon reflection and transmission. Both the amplitude and phase of $r_i$ and $t_i$ are functions of the frequency of light. For a linear cavity setup like Fig.1, the input field ($E_{in}$) of the cavity simply relates to the output ($E_t$) through $E_t/E_{in} = |t_1||t_2|exp\left[j\left(\phi_{t_1}+\phi_{t_2}\right)\right]/\left(1-|r_1||r_2|exp\left[j\left(\phi_{r_1}+\phi_{r_2}\right)\right]exp(2jn_0\omega L/c)\right)$, where $\omega$ denotes the angular frequency of the optical field. Consequently, the cavity transmission, $T_c \equiv |E_t/E_{in}|^2$, is

$$T_c = \frac{|t_1|^2 |t_2|^2}{1+|r_1|^2 |r_2|^2 - 2|r_1||r_2|\cos(\Phi_{total})} \quad (1)$$

where $\Phi_{total} = \phi_{r_1} + \phi_{r_2} + \phi_c$. $\phi_c = 2n_0\omega L/c$ denotes the phase accumulated through propagation inside the cavity. The medium in the cavity is assumed to be dispersion-free, i.e., $n_0$ is frequency-independent. Thus $\phi_c$ is only proportional to the frequency of the light. When $\Phi_{total}(\omega_0) = 2m\pi$ ($m$: positive integer), the cavity is known as on resonance at frequency $\omega_0$. When the frequency shifts to $\omega_0 + \Delta\omega$, the total phase change of $\Phi_{total}(\omega_0)$ is given by:

$$\Delta\Phi_{total} = \Phi_{total}(\omega_0 + \Delta\omega) - \Phi_{total}(\omega_0) = \phi_{r_1}(\omega_0 + \Delta\omega) - \phi_{r_1}(\omega_0) + \phi_{r_2}(\omega_0 + \Delta\omega) - \phi_{r_2}(\omega_0) + \frac{2n_0\Delta\omega L}{c}$$

(2)

Without losing the generosity, we can assume the two reflectors to be identical (i.e., $\phi_{r_1} = \phi_{r_2} = \phi_r$ and $\phi_{t_1} = \phi_{t_2} = \phi_t$). Thus,

$$\Delta\Phi_{total} = 2\left(\phi_r(\omega_0 + \Delta\omega) - \phi_r(\omega_0) + \frac{n_0\Delta\omega L}{c}\right) \simeq \frac{2L\Delta\omega}{c}\left(n_0 + \frac{c\phi'_r}{L}\right) \quad (3)$$

where $\phi'_r = \partial\phi_r/\partial\omega\big|_{\omega=\omega_0}$ is the first-order derivative term in Taylor expansion of $\phi_r(\omega,\xi)$ around $\omega_0$. This approximation holds since $\Delta\omega \ll \omega_0$ in this case. As seen from eqn. (3), when $\phi'_r = -\frac{n_0 L}{c}$, $\Delta\Phi_{total}$ would vanish and $\Phi_{total}(\omega_0 + \Delta\omega)$ would remain $2m\pi$. If this is ensured for a relatively wide range of $\Delta\omega$, which means light of different frequencies are resonant simultaneously with the cavity, the so-called WLC condition is achieved. Physically, $\phi'_r$, determined from the relative phase of the individual component of the grating response, represents the time difference between the arrival of the frequency components, and is thus called the group delay. A negative group delay means that high frequency travels a shorter distance in the reflector than the lower frequency, cancelling out the extra phase gained in cavity propagation. As such, after a round trip, the high and low frequencies both have the increase in phase for $2m\pi$ inside the WLC.

## 3. Theoretical Analysis of Linearly Chirped Bragg Grating (LCBG)

As illustrated in section 2, the ability to manipulate the penetration depth of light inside the reflector according to its wavelength is crucial for realizing a WLC. The negative group delay demands light with short wavelength to leave the grating earlier than the longer wavelength. If each wavelength can be associated with a desired reflection point along the length of the reflector, then the reflector will be considered an ideal candidate of the reflector. As is well-known, in a uniform grating, when the grating period equals to an integer multiple of the wavelength (the so-called Bragg condition), the reflected waves constructively interfere with each other, resulting in high total reflection close to unity. A LCBG has a Bragg condition varying as a function of position along the grating. This is achieved by ensuring that the spatial frequency of the grating, $\kappa$, varies as a function of position along the FBG. The Bragg condition for the chirp FBGs can be written as $k(z) = m\frac{\kappa(z)}{n_{eff}}$ where z is the position along the grating. With this type of structure, the device becomes broadband in response with a varying Bragg condition along the distance. In essence this results in a device which reflects the varying frequency components of the pulse at different points along its length as shown in Fig. 2. Consequently, the long wavelength components of the pulse will see a different delay to the short wavelength components. If the profile of the delay exactly cancels out the phase difference accumulated in the cavity propagation, the WLC condition is fully reached.

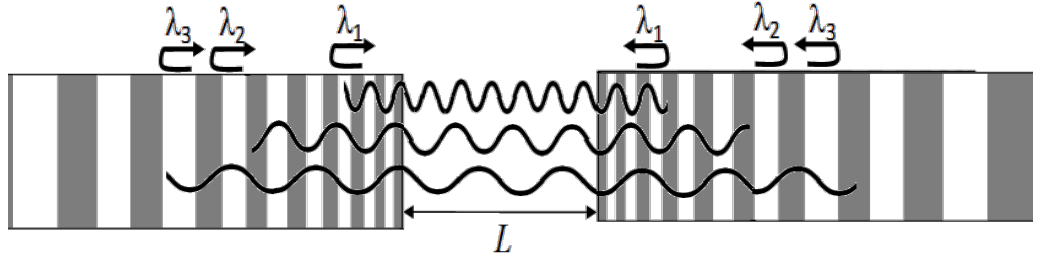

Fig.2. Schematic illustration of a typical Fabry-Perot (FP) cavity of length L formed by a pair of LCBGs. Different wavelengths ($\lambda_1<\lambda_2<\lambda_3$) are reflected at different locations inside the grating region

In order to analyze the properties of the LCBG quantitatively, we approach the reflection and transmission of the LCBG with the well-known coupled-mode theory introduced by Kogelnik.

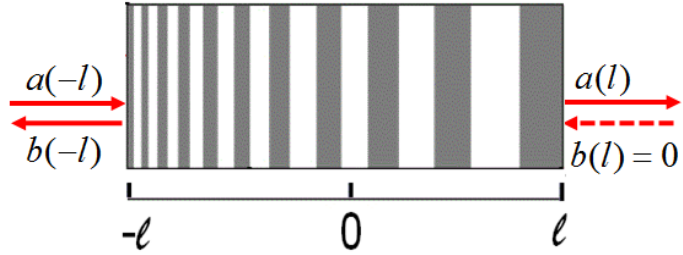

Fig.3. Coupled-mode model of light propagating in the LCBG

As shown in Fig. 3, the total field, represented by the superposition of the forward and backward propagating modes, is written as $E_{total}(z)=E_f(z)+E_b(z)=a(z)e^{ikz}+b(z)e^{-ikz}$. Thus the reflection coefficient at the entrance of the LCBG is defined as $r=\dfrac{E_b(-L)}{E_f(-L)}=\dfrac{b(-L)}{a(-L)}e^{i2kL}$

The index modulation of the LCBG along the z-axis can be represented by $\delta n(z)=\beta\cos[2\theta(z)]$ where $\beta$ is the modulation depth. $\theta(z)=\alpha z^2/2+\kappa$ in which $\kappa$ denotes the modulation frequency at $z=0$ and $\alpha$ represents the chirping parameter. For an LCBG of length $2l$ with the origin at the center, the analytical solution of $r$, solved from the coupled-mode equation, is found to be

$$r=\frac{exp(j\alpha z_0^2/2)}{jk_0\beta}\frac{F\left(j\eta;1/2;-j\alpha(l/2+z_0)^2/2\right)+\beta^2 k_0^2\rho(-l/2-z_0)F\left(1/2+j\eta;3/2;-j\alpha(l/2+z_0)^2/2\right)}{\rho F\left(-j\eta;1/2;j\alpha(l/2+z_0)^2/2\right)+(-l/2-z_0)F\left(1/2-j\eta;3/2;j\alpha(l/2+z_0)^2/2\right)}e^{-i2kL}$$

,

$$\rho=-\frac{F\left(j\eta;1/2;j\alpha(l/2-z_0)^2/2\right)}{\beta k_0^2(l/2+z_0)F\left(j\eta;3/2;j\alpha(l/2+z_0)^2/2\right)}$$

(4)

where $F(a;b;x)=\sum_{n=0}^{\infty}\dfrac{a_n}{b_n}\dfrac{x^n}{n!}$; $a_0=1$; $b_0=1$; $a_n=a(a-1)\cdots(a-n+1)$; $b_n=b(b-1)\cdots(b-n+1)$, is the

first kind of the hypergeometrical confluent function. $z_0 = (2k - \kappa)/\alpha$ and $\eta = \beta^2 k_0^2/(2\alpha)$ are expressed in terms of $k_0 \equiv \kappa/2$. According to Shapiro [14], $z_0$ implies the point where a wave of the wave number $k$ propagating along z-axis is reflected. Thus $z_0$ is relevant to the effective optical length for that particular frequency: $l_{eff} = z_0(\omega) + \ell/2$. We calculated amplitude and phase of $r$ for a negatively chirped grating (spatial frequency decreases with distance) in Fig. 3 where $2\ell = 1 \times 10^{-4}$ m, $\alpha = -6 \times 10^6 \text{m}^{-1}$ and $\beta_0 = 6.7 \times 10^{-4}$, as shown in Fig. 4

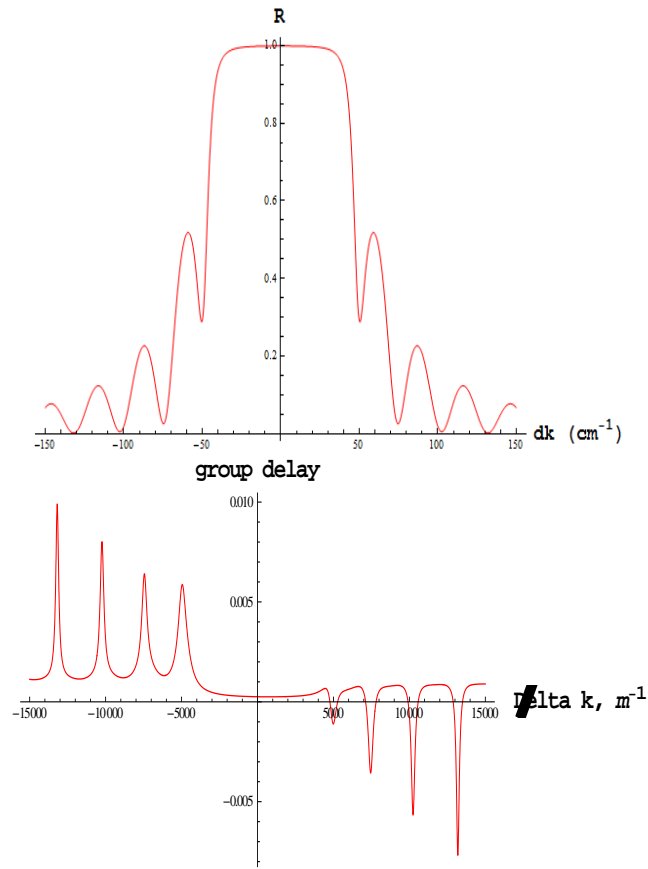

Fig.4. The theoretical amplitude and phase of reflection coefficient for a negatively chirped LCBG.

As seen from Fig.4, to our surprise, instead of being a negative value near the resonance as predicted, $\phi'_r > 0$ for the detuning region near resonance. We have selected different parameters for modulation depth, chirping rate and grating length, however, the group delay near resonance remains positive for all the combinations.

**4. Simulating the LCBG with Transfer Matrix Method (TMM)**

To verify the theoretical analysis, we simulate the reflection properties of the LCBG with numerical method. The transfer matrix method (TMM) is the most favored technique for modeling LCBG because of its speed, accuracy and the ability to handle gratings with arbitrary profiles. For this method, the grating is divided into slices much smaller in length than the grating period, so the LCBG may be considered as a piecewise representation of the continuous whole. The index change is taken to be constant within each section. Therefore, rather than a solution derived from the whole index profile of the LCBG, a discretized solution may be sought to solve each slice of the structure in a piecewise fashion, the product of these solutions approximating to the real solution of the grating structure. The detailed mathematic representation of the TTM can be found in ref [14]. Fig. 5 shows the simulation result for amplitude and phase of *r* for the negative LCBG analyzed in section 3. Excellent agreement is found with the theoretical analysis in Fig. 4. Note that the group delay near the resonance frequency, calculated form numerical simulation, is also positive.

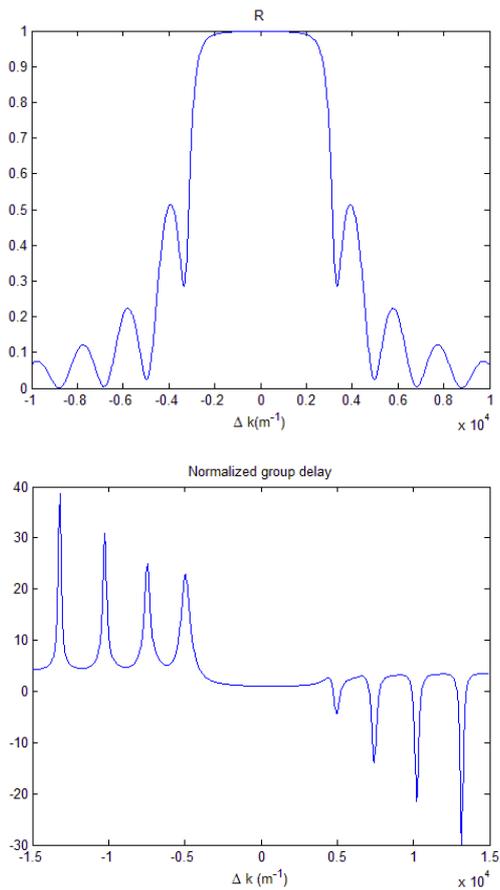

Fig.5. The simulated result for amplitude and phase of reflection coefficient for the LCBG.

**5. Revisiting the reflection properties of the LCBG**

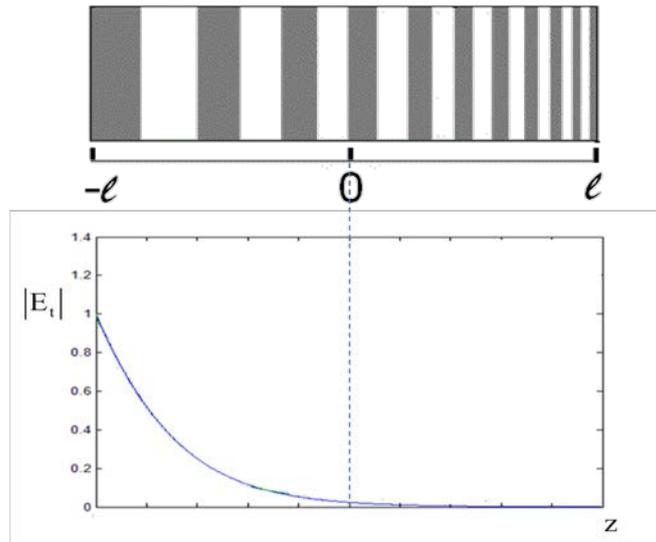

Fig.6. The attenuation of forward-propagating wave in the LCBG.

The discrepancy between the physical intuition about the reflection in the LCBG and the actual analysis rooted in the ignorance of the wide band gap of the LCBG. Effectively, we can treat the LCBG as a concatenation of individual uniform FBG's, each occupying a center reflection wavelength. The physical intuition simply assumes that light is only reflected at the piece of the LCBG where Bragg condition is matched, while the other pieces in the grating have trivial effect on the reflection. However, the periodical structure of uniform gratings generates a wide reflection band near the resonance. A closer look at the field profile of the forward-propagating wave in the LCBG, as shown in Fig. 6, reveal that even before reaching the Bragg condition point, a significant portion of the incident wave has already been reflected. Therefore, the reflection by other pieces inside the LCBG, even where the Bragg condition is not met, should not be neglected.

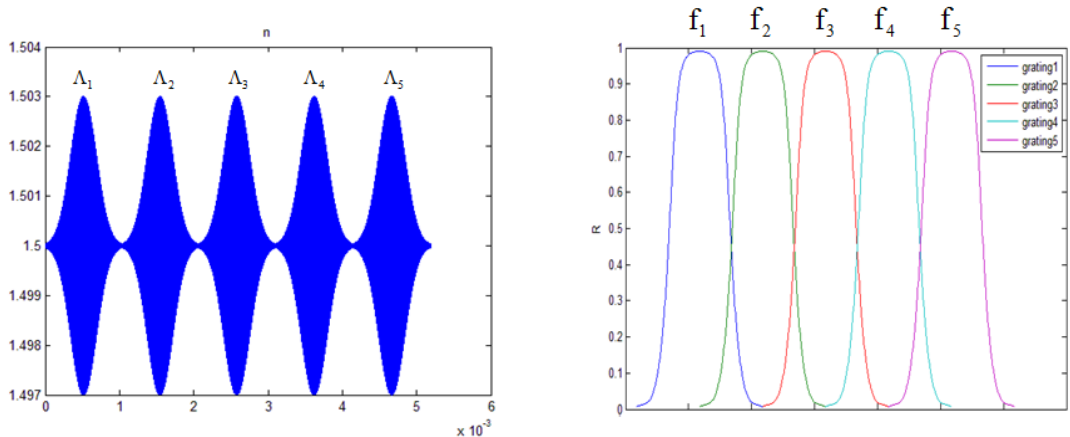

Fig.7. The scheme and reflection spectrum of segmented apodized chirped grating.

To illustrate this point, we further simulated a segmented apodized chirped grating, with the center Bragg wavelength of individual gratings deliberately set far from each other so that their reflection spectra don't overlap with each other, as shown in Fig. 7. In this case, the light resonating with one grating is not affected by other gratings. Reflection doesn't occur until the Bragg match condition is reached. By carefully design the grating we can achieve quasi-negative-group-delay between the individual frequencies, and therefore simultaneous resonance as shown in Fig. 8 (a) and (b). However, this scheme is only valid for discrete and limited number of frequencies. When the frequency is off resonance, the phase of reflection is purely random as shown in Fig. 8 (c) and (d), resulting a random group delay. In the case of the LCBG, due to the continuous structure, the effect of scattering by the off-resonance gratings is unavoidable. As such, it is improper to claim that the penetration depth of the light in the LCBG is linearly related to its wavelength. The simple LCBG is not suitable for constructing WLC. Whether it is possible to engineer the grating in other aspects to realize the negative group delay and thus the WLC is still an open question.

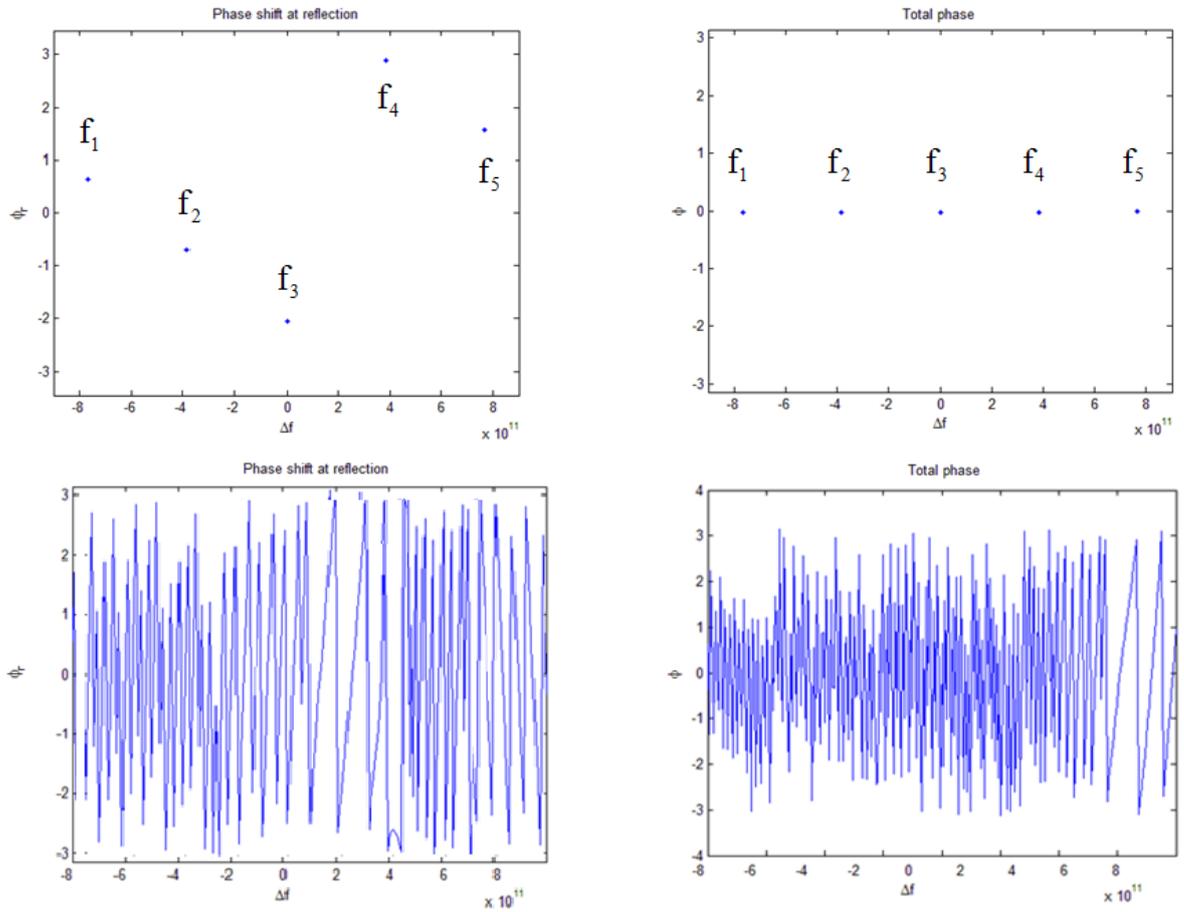

Fig.8. The total phase for a round-trip propagation in the cavity based on segmented apodized chirped grating.

## 6. Conclusion

In conclusion, we studied the possibility of making use of such a dispersion-compensating property to create a white light cavity (WLC). A pair of LCBGs is suggested as the reflectors of the cavity. The analytical and numerical analysis show that the reflection by a LCBG does not solely occurs at the Bragg matching point. The accumulated effect of multiple scatterings at different location inside the LCBG produces a positive group delay, which is contradictory to conventional notion about the LCBG.